\documentclass{osa-article}
\global\arraycolsep=2pt
\journal{osajournal}

\articletype{Research Article}

\begin{document}

\title{Casimir-Polder forces in inhomogeneous backgrounds}

\author{Kimball A. Milton\authormark{1}\authormark{*}}

\address{\authormark{1} Homer L. Dodge Department of Physics and
Astronomy, University of Oklahoma, Norman, OK 73019 USA}

\email{\authormark{*}kmilton@ou.edu} 

\homepage{http://nhn.ou.edu/$\sim$milton} 


\begin{abstract}
Casimir-Polder interactions are considered in an inhomogeneous, dispersive
background.  We consider both the interaction between a polarizable atom
and a perfectly conducting wall, and between such an atom and a plane
interface between two different inhomogeneous dielectric media.  
Renormalization is
achieved by subtracting the interaction with the local inhomogeneous
medium by itself.  The results are expressed in general form, and 
generalize the Dzyaloshinskii-Lifshitz-Pitaevskii interaction between a
spatially homogeneous dielectric interface and a polarizable atom.
\end{abstract}

\section{Introduction}
In 1948 Casimir and Polder \cite{Casimir:1947hx} first incorporated 
retardation, the finite
speed of light, into the quantum interaction between neutral polarizable
atoms, thus generalizing the van der Waals interaction worked out by
London in 1930 \cite{london1930}.  
They considered both the interaction between two such
atoms, and the interaction of such an atom with a perfectly conducting
wall. In both cases, the background was considered to be vacuum.  Both 
types of interactions are referred to as Casimir-Polder (CP) interactions.
In this paper we consider the CP interaction 
in an inhomogeneous medium between an atom and a
flat interface, which might be a perfectly conducting plate or a plane
across which the permittivity is nonanalytic.  For reference, we remind
the reader of the CP potential between an atom having isotropic electric
polarizability $\alpha$ and a perfect mirror at a distance $Z$ away in
vacuum
($\hbar=c=1$):
\begin{equation}
    E_{\rm CP}=-\frac{3\alpha}{8\pi Z^4},
\end{equation}
where the transverse magnetic (TM) and transverse electric (TE) 
contributions are in the ratio $5:1$.  This has been observed in
experiments \cite{hinds,cornell}.

This paper is part of a program of investigating Casimir energies and 
stresses in inhomogeneous media. Casimir forces themselves were 
originally derived for bodies separated by vacuum by Lifshitz 
\cite{lifshitz}, and only later generalized to a background consisting
of a homogeneous dielectric \cite{dlp}.  Relatively little has been
done when the background medium is inhomogeneous.  In the Casimir
context, we can cite the work of Griniasty and Leonhardt
\cite{griniasty1,griniasty2}, and of Bao et al.~\cite{bao1,bao2}. 
Our group had first investigated scalar fields
in an inhomogeneous half-space, studying the divergences that arise, as
well as the singularities that occur in the vacuum expectation value
of the stress tensor and the energy density \cite{Milton:2016sev},
and earlier references cited therein.  Efforts were made to
effect a suitable renormalization \cite{Fulling:2018qcn}. 
We generalized these investigations to
electromagnetism, and found universal behaviors for both divergences
and singularities \cite{Parashar:2018pds}. 
Most germane to the present investigation is the
computation of energies and stresses between perfect reflectors, and 
between 
parallel planes of nonanlyticity, when the media exhibit both spatial 
variation
(in a single perpendicular coordinate, $z$), and dispersion \cite{ihm}.

In the next section we will consider an atom above a conducting plane,
but now with a medium which has a permittivity $\varepsilon(z,\omega)$ depending both on the position
$z$ above the plane, and on the frequency $\omega$, that is, the medium
may be dispersive.  (We will for simplicity disregard dissipation.)
We will usually suppress the dependence on frequency in our formulas.
In Sec.~\ref{sec3} we will generalize this situation to two parallel
dielectrics, one with permittivity $\varepsilon_1(z)$ and the
second having permittivity $\varepsilon_2(z)$, the first extending
from $-\infty<z<0$ and the second from $0<z<\infty$; we assume the atom
is in the positive $z$ region. We assume that $\varepsilon_{1,2}$ are
analytic functions of $z$ in their respective regions.
The expressions directly obtained are
divergent; they will be renormalized by subtracting the energy that would
be present if only one inhomogeneous medium would be present, i.e., in the
absence of the wall or discontinuity.  The resulting expressions are finite
and generalize the usual CP interaction energy.  Sec.~\ref{sec4} gives
numerical results for a nontrivial solvable model.  Concluding remarks
are offered in Sec.~\ref{sec5}.

\section{CP interaction with perfectly conducting wall}\label{sec2}
For the planar geometry considered, the Green's function describing the
background can be decomposed into two parts, TE and TM. The corresponding
reduced Green's functions are denoted by $g^{E,H}$. The CP interaction
energy between the background and a polarizable atom can be written 
in terms of the trace of the electromagnetic Green's dyadic, or as
(see, for example, Ref.~\cite{Parashar:2018pds})
\begin{eqnarray}
    E_{\rm CP}&=&-\frac12\mbox{Tr} \,\boldsymbol{\Gamma}(Z,Z)4\pi \boldsymbol{\alpha}(Z)\nonumber\\
    &=&-2\pi\alpha\int_{-\infty}^\infty 
    \frac{d\zeta}{2\pi}\int\frac{(d\mathbf{k}_\perp)}{(2\pi)^2}
    \left[\frac1{\varepsilon(z)^2}\partial_z\partial_{z'}g^H(z,z')
    +\frac{k^2}{\varepsilon(z)^2}g^H(z,z')-\zeta^2g^E(z,z')\right]_{z'=z}.
    \label{cpenergy}
\end{eqnarray}
Here we have made a Euclidean rotation to imaginary frequencies,
$\omega\to i\zeta$, and the dependence on $\zeta$ of the permittivity
and Green's functions is implicit.
We will in this section consider a perfectly conducting boundary at $z=0$,
while the atom is located at $z=Z>0$.
\subsection{TE mode}
The TE Green's function satisfies
\begin{equation}
    [-\partial_z^2+k^2+\zeta^2\varepsilon(z)]g^E(z,z')=\delta(z-z'),
\end{equation}
which can be solved in terms of solutions of the corresponding homogeneous
equation, $F$ and $G$, where $F$ goes to zero as $z\to\infty$ and $G(0)=0$.
The construction of the Green's function is
\begin{equation}
    g^E(z,z')=\frac{F(z_>)G(z_<)}\beta,\label{ge}
\end{equation}
where $\beta$ is the Wronskian, $\beta=[F,G]$, a constant in this case,
and where we introduced the notation
\begin{equation}
    [f,g](z)=f(z)g'(z)-f'(z)g(z),\label{wronskian}
\end{equation}
which notation will be used even if $f$ and $g$ do not satisfy the same
differential equation.

Inserting this construction into the formula for the TE energy 
(\ref{cpenergy}) yields a divergent expression; even if the reflecting
wall were not present, the atom would experience an infinite force
from the inhomogeneous medium.  We subtract the energy due to filling
all of space for $z<0$ with the (unique) analytic continuation of 
$\varepsilon(z)$, imposing a convenient boundary condition at the
nearest singularity on the real axis.  This singularity may be at
$-\infty$.  In most cases it will suffice to choose the solution that
vanishes at the singularity.  
Doing so corresponds to a Green's function of the same
form as Eq.~(\ref{ge}) except that $G$ is replaced by $\tilde G$, which
satisfies the same differential equation but vanishes not at zero but at
the singular point, symbolically denoted $-\infty$. 
(We will give an explicit example of the singularity occurring  at a finite
point in Sec.~\ref{sec4}.)
If we normalize $\tilde G$ to retain 
the same Wronskian in the two situations, we see that
\begin{equation}
    G(z)=\tilde G(z)-\frac{\tilde G(0)}{F(0)}F(z).
\end{equation}
If we subtract the reference energy (without the plate) from the energy
with the plate, we obtain the finite ``renormalized'' energy, which 
refers to the interaction between the plate and the atom:
\begin{equation}
E^{\rm TE}_R=-2\pi\alpha\int\frac{d\zeta (d\mathbf{k}_\perp)}{(2\pi)^3}
\zeta^2\frac{\tilde G(0)}{F(0)}\frac{F(Z)^2}{\beta}.\label{tepc}
\end{equation}

A small check of this formula is to examine what happens
when the medium is a vacuum, so
\begin{equation}
    F(z)=e^{-\kappa z},\quad \tilde G(z)=e^{\kappa z},\quad \beta=2\kappa,
    \label{vacfg}
\end{equation}
and then we immediately find the expected TE CP energy:
\begin{equation}
    E^{\rm TE}_{\rm CP}=-\frac\alpha{16\pi Z^4}.\label{vactepc}
\end{equation}

\subsection{TM mode}
The TM Green's function obeys
\begin{equation}
    \left(-\partial_z\frac1{\varepsilon(z)}\partial_z+\frac{k^2}
    {\varepsilon(z)}    +\zeta^2\right)g^H(z,z')=\delta(z-z'),
\end{equation}
which again may be solved by the construction (\ref{ge})
where now $F$ and $G$ are solutions of the homogeneous equation whose
derivatives vanish at $z=\infty$ and $z=0$, respectively.  Then, following
the procedure above we find the renormalized TM contribution to the CP
interaction energy:
\begin{equation}
        E^{\rm TM}_R=2\pi \alpha \int_{-\infty}^\infty 
        \frac{d\zeta}{2\pi}\int\frac{(d\mathbf{k}_\perp)}{(2\pi)^2}
        \frac1{\beta\varepsilon(z)^2}\frac{\tilde G'(0)}{F'(0)}\left[
        F'(Z)^2+k^2F(Z)^2\right].\label{pctm}
\end{equation}
where $\tilde G$ is the solution to the homogeneous equation with
the boundary condition of vanishing at $-\infty$, when the inhomogeneous
dielectric extends over all space.  In this case, the Wronskian is
not constant, but is related to $\beta$ by $W^H(z)=\beta^H\varepsilon(z)$.
Again, we can check this result
by specializing to a vacuum medium, as in Eq.~(\ref{vacfg}), which yields the expected result
\begin{equation}
    E^{\rm TM}_{CP}=-\frac{5\alpha}{16\pi Z^4}.\label{vactmcp}
\end{equation}

\section{Dielectric half-spaces}
\label{sec3}
Now we consider a background consisting of two analytic permittivities,
\begin{equation}
    \varepsilon(z)=\left\{\begin{array}{cc}
      \varepsilon_1(z),   &z<0,  \\
        \varepsilon_2(z), &z>0,
    \end{array}\right.
\end{equation}
For the TE mode, we can, of course, write the reduced Green's function
as
\begin{equation}
    g^E(z,z')=\frac1\beta \mathcal{F}(z_>)\mathcal{G}(z_<),
    \label{gfconst}
\end{equation}
where $\beta$ is a constant, in terms of solutions of the equation
\begin{equation}
\left[-\partial_z^2+k^2+\zeta^2\varepsilon(z)\right]\left\{\begin{array}{c}
     \mathcal{F}  \\
     \mathcal{G}
\end{array}=0,\right.
\end{equation}
where $\mathcal{F}\to0$ as $z\to+\infty$, $\mathcal{G}\to 0$ as 
$z\to-\infty$. Let the fundamental solutions in the two different
regions of analyticity be denoted $F_1, G_1$ and $ F_2, G_2$, 
respectively, with the boundary conditions that $F_2\to 0$ as 
$z\to+\infty$,
$G_1\to 0$ as $z\to-\infty$. Then, requiring continuity of 
$\mathcal{F}$ and $\mathcal{G}$ and their derivatives leads to
($z,z'>0$)
\begin{equation}
    g^E(z,z')=\frac{1}{\beta\beta_2}F_2(z_>)\left\{[F_2, G_1](0)G_2(z_<)
-[G_2, G_1](0)F_2(z_<)\right\},\label{gedb}
\end{equation}
where we have used the Wronskian notation (\ref{wronskian}), while
the (constant) Wronskians are 
\begin{equation}
    \beta=[\mathcal{F},\mathcal{G}],\quad \beta_2=[F_2, G_2].
\end{equation}
$\beta_2$ is arbitrary, depending on how the functions are
normalized, but it follows from Eq.~(\ref{gedb}) that $\beta=[F_2, G_1](0)$.
Therefore, the CP energy of the atom is given by
\begin{equation}
    E^{\rm TE}=-\frac{2\pi\alpha}{\beta_2}
    \int_{\infty}^\infty\frac{d\zeta}{2\pi}
    \int\frac{(d\mathbf{k}_\perp)}{(2\pi)^2}(-\zeta^2)F_2(Z)\left(
    G_2(Z)-\frac{[G_2, G_1](0)}{[F_2, G_1](0)}F_2(Z)\right).
\end{equation}
This expression will be divergent, because it includes the interaction
with the background dielectric without the interface being present.
Renormalization here consists of subtracting the corresponding expression
where the dielectric 2 extends over all space.  This corresponds to the
reference energy
\begin{equation}
    E^{\rm TE}_{\rm ref}=-2\pi\alpha\int\frac{d\zeta(d\mathbf{k}_\perp)}
    {(2\pi)^3}(-\zeta^2)\frac{F_2(Z) G_2(Z)}{\beta_2},
\end{equation}
where now we further assume $G_2\to 0$ as $z\to -\infty$. (We can choose
the same $G_2$ in the original configuration.) 
(The same caveat mentioned above applies if $\varepsilon_2$ develops
a singularity in region 1.)
Subtracting this from
the original energy gives the renormalized interaction energy between
the interface and the atom:
\begin{equation}
    E_R^{\rm TE}=-\frac{2\pi\alpha}{\beta_2}\int\frac{d\zeta 
    (d\mathbf{k}_\perp)}{(2\pi)^3}\zeta^2
    \frac{[G_2, G_1](0)}{[F_2, G_1](0)}F_2(Z)^2.
\end{equation}
It is easy to check that this reduces to the result (\ref{tepc}) when
medium 1 is replaced by a uniform dielectric with $\varepsilon\to\infty$.

The same procedure applies to the TM mode.  The construction of the 
Green's function is the same as in Eq.~(\ref{gfconst}) but where now
the functions satisfy
\begin{equation}
    \left(-\partial_z\frac1{\varepsilon(z)}\partial_z+\frac{k^2}
    {\varepsilon(z)}+\zeta^2\right)\left\{\begin{array}{c}
        \mathcal{F}  \\
        \mathcal{G} 
    \end{array}\right.=0,
\end{equation}
and the Wronskian is $W(z)=\beta\varepsilon(z)$.  Correspondingly,
the generalized Wronskian is defined as
\begin{equation}
    [f_i,g_j]_\varepsilon(z)=f_i(z)\frac{g'_j(z)}{\varepsilon_j(z)}-
    g_j(z)\frac{f'_i(z)}{\varepsilon_i(z)},
\end{equation}
where, as above, the subscripts refer to the two regions 1 and 2.
Now it is $\mathcal{F}$, $\mathcal{G}$ and $\mathcal{F}'/\varepsilon$,
$\mathcal{G}'/\varepsilon$ which are continuous.  Then following the
same procedure as for the TE case, we find for the renormalized (background
subtracted) TM energy
\begin{equation}
    E^{\rm TM}_R=\frac{2\pi\alpha}{\beta_2}\int\frac{d\zeta (d
    \mathbf{k}_\perp)}{(2\pi)^3}\frac{1}{\varepsilon_2(Z)^2}
    \frac{[G_2, G_1]_\varepsilon(0)}{[F_2, 
    G_1]_\varepsilon(0)}\left[F_1'(Z)^2+k^2F_1(Z)^2\right].
\end{equation}
This is easily seen to reduce to the corresponding perfect conductor
energy (\ref{pctm}) when the first medium has infinite permittivity.

\section{Exactly solvable model}
\label{sec4}
\begin{figure}[h!]
\centering\includegraphics[width=7cm]{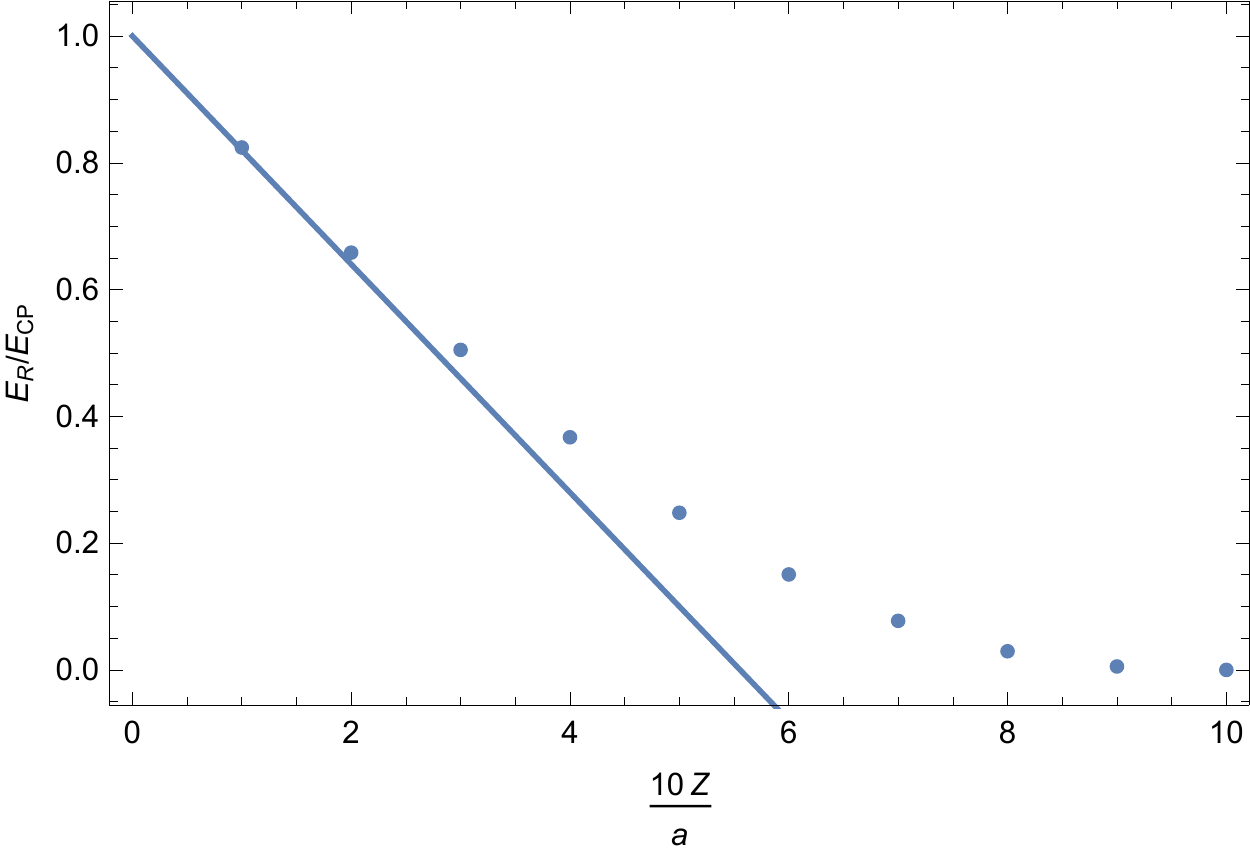}
\caption{\label{fig1} The renormalized TE CP energy for the potential (\ref{quadpot}),
describing the medium bwtween a polarizable atom and a perfectly
conducting plate, relative to the vacuum value (\ref{vactepc}). The dots are the 
results of numerical integration, while the straight line is 
the perturbative approximation (\ref{pte}).}
\end{figure}

There are only a few cases where both the TE and TM equations may be
solved in terms of known functions.  One of them is the potential
\cite{griniasty2,Parashar:2018pds} 
\begin{equation}
    \varepsilon(z)=\frac\lambda{(a-z)^2},\label{quadpot}
\end{equation}
where now the universe is supposed to be the region $z<a$.
The fundamental solutions are
\begin{subequations}
\begin{align}
F^{E,H}&=(a-z)^{\pm1/2}I_\nu(k(a-z)),\\
\tilde G^{E,H}&=(a-z)^{\pm1/2}K_\nu(k(a-z)),
\end{align}
\end{subequations}
with $\nu=\sqrt{\lambda\zeta^2+1/4}$. 
The corresponding effective Wronskians are $\beta^E=1$, $\beta^H=1/\lambda$. As an illustration, in 
Fig.~\ref{fig1} we plot the resulting renormalized TE CP energy 
(\ref{tepc}) for an atom above a plate in such a medium, for various values
of $Z/a$, relative to the vacuum CP energy (\ref{vactepc}), for the
special case when $\lambda/a^2=1$, so the permittivity approaches unity
at the plate.
This is compared with the perturbative approximation, as described in
Ref.~\cite{Parashar:2018pds}, which is readily worked out to be
\begin{equation}
    E^{\rm TE}_p=-\frac{\alpha}{16\pi Z^4}\left(1-\frac{9}5 Z\right).
    \label{pte}
\end{equation}
Note that as expected, when the atom is close to the plate,
the interaction coincides with the vacuum value, but screening sets in
at larger distances.

\begin{figure}[h!]
\centering\includegraphics[width=7cm]{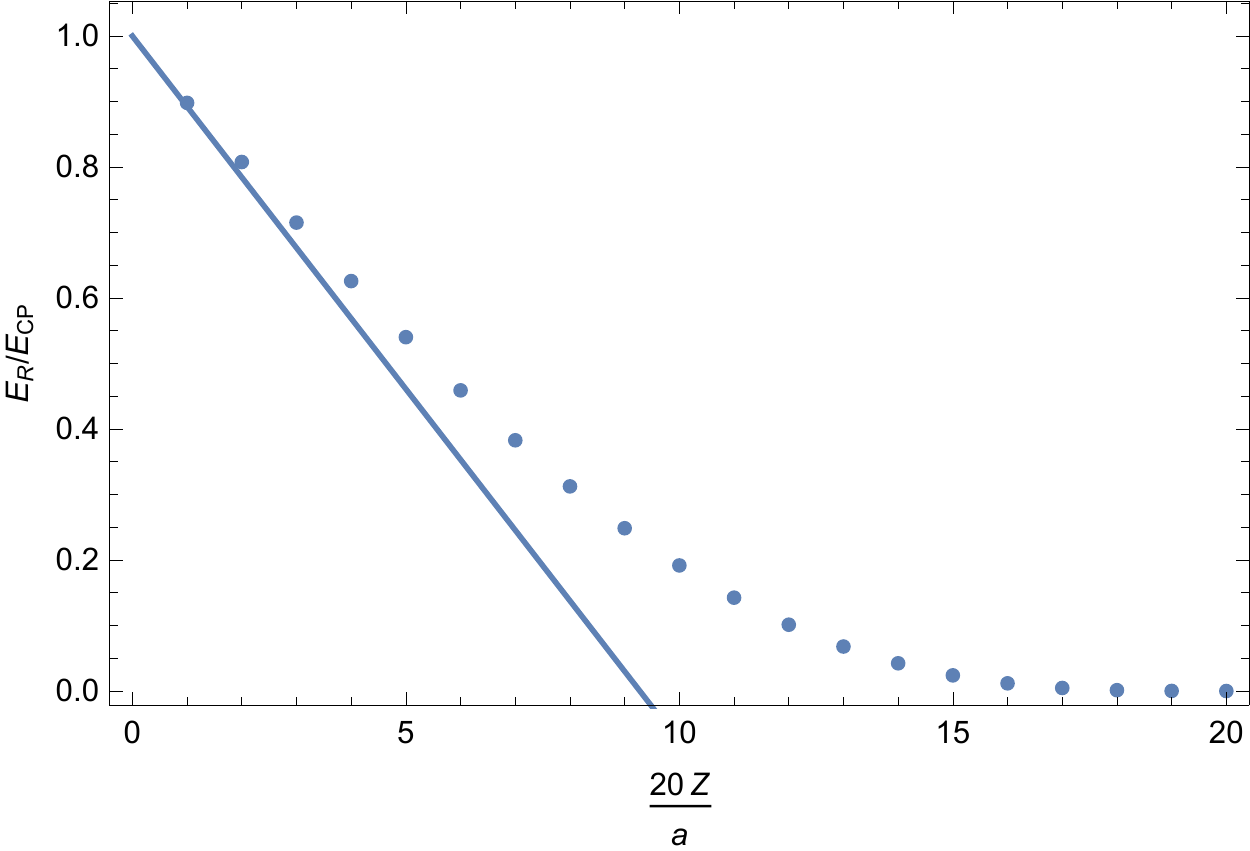}
\caption{\label{fig2} The renormalized TM CP energy for the potential 
(\ref{quadpot}),
describing the medium between a polarizable atom and a perfectly
conducting plate, relative to the vacuum value (\ref{vactmcp}). 
The dots are the 
results of numerical integration, while the straight line is the
perturbative approximation (\ref{ptm}).}
\end{figure}

The corresponding results for the TM mode are shown in Fig.~\ref{fig2}.
Note now the perturbative approximation is 
\begin{equation}
    E^{\rm TM}_p=-\frac{5\alpha}{16\pi Z^4}\left(1-\frac{97}{45} Z\right).
    \label{ptm}
\end{equation}

\begin{figure}[h!]
\centering\includegraphics[width=7cm]{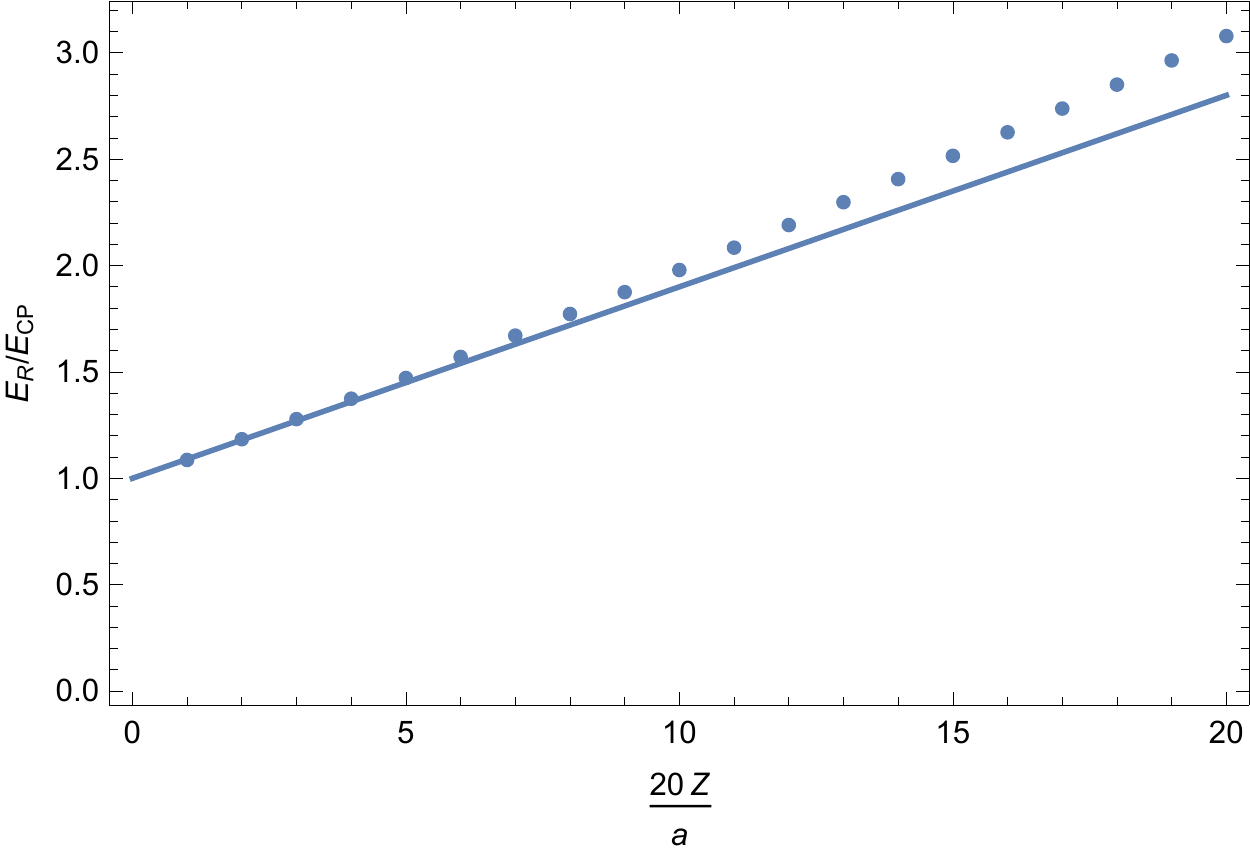}
\caption{\label{fig3} The renormalized TE CP energy for the potential 
(\ref{quadpot2}),
describing the medium between a polarizable atom and a perfectly
conducting plate, relative to the vacuum value (\ref{vactepc}). 
The dots are the 
results of numerical integration, while the straight line is 
the relative perturbative
correction $1+\frac95 Z$.  It is seen this is the continuation of the
results seen in Fig.~\ref{fig1}.}
\end{figure}

Now we illustrate what can happen when the analytically continued
permittivity develops a singularity in the extended region.  So consider
the permittivity
\begin{equation}
    \varepsilon(z)=\frac\lambda{(a+z)^2},\label{quadpot2}
\end{equation}
which now has the singularity at $-a$.  The solutions are like before,
with $(a-z)$ replaced by $(a+z)$ but now we have to interchange the
roles of $I_\nu$ and $K_\nu$ because we want $F$ to vanish at infinity,
and $\tilde G$ to be well-behaved at $z=-a$.  The formula for the
renormalized TE energy is then
\begin{equation}
    E_R^E=-\frac\alpha{\lambda^{3/2}\pi}(a+Z)\int_0^\infty dk\,k\int_{1/4}
    ^\infty d\nu\,\nu\sqrt{\nu^2-1/4}\frac{I_\nu(ka)}{K_\nu(ka)} 
    K^2_\nu(k(a+Z)),
\end{equation}
where we have changed the integration variable from $\zeta$ to $\nu$.
Again we consider the case $\lambda/a^2=1$.
The perturbative approximation to this is Eq.~(\ref{pte}) with the reversed
sign of the linear term.  The numerical integration is compared with the 
perturbative approximation in Fig.~\ref{fig3}.

\section{Conclusion}
\label{sec5}
We have show how to isolate a finite, ``renormalized'' Casimir-Polder 
interaction
energy between a plane surface of nonanalyticity, which includes a
perfect reflector, and an isotropic polarizable atom, separated by an
inhomogeneous dispersive medium. (For simplicity the inhomogeneity is 
restricted to a single direction, perpendicular to the plane interface.)
Incorporation of anisotropy and of dispersion in the polarizability is 
readily achievable. 
Elsewhere we will consider the more complicated problem of the
Casimir interaction between two such interfaces \cite{ihm}.

\section*{Funding}

National Science Foundation (NSF) (1707511)
\\

\section*{Acknowledgments}
I thank my collaborators Li Yang, Prachi Parashar, Hannah Day, Gerard
Kennedy, Pushpa Kalauni, In\'es Cavero-Pel\'aez, and Steve Fulling.





\end{document}